\documentclass[prl,twocolumn,amsmath,amssymb]{revtex4}
\usepackage[latin2]{inputenc}
\usepackage{amsmath}
\usepackage{graphicx}
\usepackage{dcolumn}
\usepackage{epstopdf}

\newcommand{\ba}{\begin{eqnarray*}}
\newcommand{\ea}{\end{eqnarray*}}
\newcommand{\baa}{\begin{eqnarray}}
\newcommand{\eaa}{\end{eqnarray}}
\def\bar{\begin{array}}
\def\ear{\end{array}}

\def\pr{^{\prime}}

\begin{document}

\title{Comment on ``Response calculations with an independent particle system with an exact one-particle density matrix''}

\author{Ryan Requist}
\email{ryan.requist@physik.uni-erlangen.de}
\author{Oleg Pankratov}
\affiliation{
Theoretische Festk\"orperphysik, Universit\"at Erlangen-N\"urnberg, Staudtstra\ss e 7-B2, 91058 Erlangen, Germany
}

\date{\today}

%\begin{abstract}

%We point out an error in the argument [PRL \textbf{105}, 013002 (2010)] that the time independence of the occupation numbers in the adiabatic approximation follows from the invariance of the ground-state interaction energy functional with respect to changes in the phases of the natural orbitals.

%\end{abstract}

\maketitle

Giesbertz, Gritsenko and Baerends (GGB) have stated that the occupation numbers $n_k$ in time-dependent density matrix functional theory are time independent in the ``adiabatic'' approximation (AA) for \textit{any} ground-state functional \cite{giesbertz2010b} (see also \cite{pernal2007a}).  It is important to know whether this statement is true as it has implications for the design of functionals capable of generating time-dependent occupation numbers.  Here we show that the argument given by GGB to support this statement is incorrect.  The statement, however, is true; it follows quite generally from the stationarity of the ground state \cite{requist2010c}.

The equation of motion for the one-body reduced density matrix $\gamma$ implies $i dn_k/dt = W_{kk}^{\dag} - W_{kk}$ \cite{pernal2007a}, where
\begin{align}
W_{kl} = \sum_{qrs} w_{kqrs} \Gamma_{srql} \label{eqn:W:matrix}
\end{align}
with $\Gamma_{srql} = \langle \Psi | \hat{c}_l^{\dag} \hat{c}_q^{\dag}\hat{c}_r\hat{c}_s | \Psi \rangle$ and 
\begin{align}
w_{kqrs} \equiv \int dx dx\pr \phi_k^*(x) \phi_q^*(x\pr) \frac{1}{|\mathbf{r}-\mathbf{r}\pr|} \phi_r(x\pr) \phi_s(x).
\end{align}  
In the AA, the memory-dependent functional $\Gamma([\gamma];t)$ on the right-hand side of Eq.~(\ref{eqn:W:matrix}) is approximated by the ground-state functional $\Gamma_0[\gamma]$ evaluated for $\gamma(t)$.  GGB argue that the invariance of the ground-state interaction energy functional $W_0=W_0[\gamma]$ with respect to the change $\phi_k \rightarrow e^{i\alpha_k} \phi_k$ in the phases of the natural orbitals implies $dn_k/dt=0$.  Therefore, they claim to prove the implication $dW_0[\gamma]/d\alpha_k=0 \Rightarrow dn_k/dt = 0$.  The crux of their argument is the statement
\begin{align}
\frac{dW_0[\gamma]}{d\alpha_k} = W_{0,kk}^{\dag} - W_{0,kk}, \label{eqn:crux}
\end{align}
where $W_{0,kk}$ are defined in the same way as the $W_{kk}$ but with ground-state quantities.  To establish Eq.~(\ref{eqn:crux}), GGB use the identity 
\begin{align}
i \frac{dW_0}{d\alpha_k} = \int dx \frac{\delta W_0}{\delta \phi_k^*(x)} \phi_k^*(x) - \int dx \frac{\delta W_0}{\delta \phi_k(x)} \phi_k(x) \label{eqn:dW}
\end{align} 
and the statement 
\begin{align}
W_{0,kk}^{\dag} = \int dx\frac{\delta W_0}{\delta \phi_k^*(x)} \phi_k^*(x), \label{eqn:Wkk}
\end{align}
quoted from Ref.~\onlinecite{pernal2005b}, where it was derived from
\begin{align}
\frac{\delta W_0[\gamma]}{\delta \phi_i^*(x)} = \sum_p \frac{\partial W_0[\gamma]}{\partial \xi_p} \frac{\delta \xi_p}{\delta \phi_i^*(x)} + \frac{1}{2} \sum_{kqrs}  \frac{\delta w_{kqrs}}{\delta \phi_i^*(x)} \Gamma_{srqk}. \label{eqn:PC1}
\end{align}
Here $W_0[\gamma]=\frac{1}{2}\mathrm{min}_{\{\xi_p\}} \sum_{kqrs} w_{kqrs} \Gamma_{srqk}(\xi_p)$ and $\{\xi_p\}$ parametrize a constrained search over $N$-representable $\Gamma_{srqk}$ that contract to $\gamma$.  It was argued \cite{pernal2005b} that Eq.~(\ref{eqn:Wkk}) follows from Eq.~(\ref{eqn:PC1}) because the first term vanishes due to the variational nature of the constrained search.  But 
there are two flaws with this argument: (i) Eq.~(\ref{eqn:PC1}) itself is manifestly incorrect because $W_0[\gamma]$ has no $\{\xi_p\}$ dependence after the constrained search has been performed: the operations $\sum_p \frac{\partial}{\partial \xi_p}$ and $\min_{\{\xi_p\}}$ do not commute; (ii) The variational character of $W_0([\phi_i];\xi_p) \equiv \frac{1}{2} \sum_{kqrs} w_{kqrs} \Gamma_{srqk}(\xi_p)$ at the minimizing $\xi_p$ for fixed $\{\phi_i, n_i\}$ does not imply that the gradient with respect to $\xi_p$ is zero, because $W_0([\phi_i];\xi_p)$ is only stationary with respect to the subspace of $\{\xi_p\}$ degrees of freedom that are orthogonal to the $\phi_i$ degrees of freedom as the latter are \textit{constrained}.  Ultimately, the argument is incorrect because it does not account for the $\phi_i$ dependence of $\Gamma_{srqk}$.

For the specific case of approximate $W_0[n_i,\phi_i,\phi_i^*]$ that contain only $w_{kqkq}$ and $w_{kqqk}$ Coulomb integrals and in which the $\Gamma_{srqk}$ are functions of $n_i$, it might seem that Eq.~(\ref{eqn:Wkk}) can be verified by an explicit calculation of the functional derivative.  However, such a calculation is not valid because the variation $\phi_k^* \rightarrow \phi_k^* + \delta \phi_k^*$ holding fixed all other $\phi_i^*$ and all $\phi_i$ corresponds to a non-Hermitian $\gamma+\delta \gamma$.  Hence, such a variation goes outside the physical domain of $W_0[n_i,\phi_i,\phi_i^*]$.  The functional derivative of $W_0[\gamma]$ with respect to an orbital should be understood as
\begin{align}
\int dx \frac{\delta W_0[\gamma]}{\delta \phi_k^*(x)} \phi_k^*(x) = n_k \int dx dx\pr \phi_k^*(x) \frac{\delta W_0[\gamma]}{\delta \gamma(x\pr,x)} \phi_k(x\pr), \label{eqn:alt}
\end{align}
where now $W_0[\gamma]=\frac{1}{2}\int dx_1 dx_2 \frac{1}{|\mathbf{r}_1-\mathbf{r}_2|} \Gamma_0([\gamma];x_1 x_2,x_1 x_2)$ and $\Gamma_0([\gamma];x_1 x_2,x_1\pr x_2\pr)$ is the ground-state two-body reduced density matrix functional \cite{gilbert1975}.  Clearly, Eqs.~(\ref{eqn:dW}) and (\ref{eqn:alt})  cannot justify Eq.~(\ref{eqn:crux}) because the right-hand side of Eq.~(\ref{eqn:crux}) depends on degrees of freedom of $\Gamma_0(x_1 x_2,x_1\pr x_2\pr)$ that are integrated out in the definition of $W_0[\gamma]$.  This information cannot be recovered by taking the derivative with respect to $\alpha_k$.

This work was supported by the Deutsche Forschungsgemeinshaft (Grant No. PA 516/7-1).

\bibliography{bibliography2010b}
\end{document}